\documentclass{elsart}

\usepackage[dvips]{graphicx}

\usepackage{amssymb}
\journal{Chemical Physics Letters}
\newcommand{\ie}{i.e.}
\newcommand{\eg}{e.g.}

\begin{document}

\begin{frontmatter}

\title{Smooth/rough layering in liquid-crystalline/gel state of dry phospholipid film, in relation to its ability to generate giant vesicles}

 \author{Mafumi Hishida},
 \author{Hideki Seto},
 \author{Kenichi Yoshikawa\corauthref{cor}}

 \corauth[cor]{Corresponding author. FAX: +81 75 753 3779}
 \ead{yoshikaw@scphys.kyoto-u.ac.jp}

 \address{Department of Physics, Graduate School of Science, Kyoto University, Kyoto 606-8502, Japan\\}

\begin{abstract}
Morphological changes in a dry phospholipid film on a solid substrate were studied below and above the main transition temperature, between the gel and liquid-crystalline phases by phase-contrast microscopy and AFM. A Phospholipid film in the liquid-crystalline phase exhibits flat, smooth layering, whereas that in the gel phase shows rough, random layering. These film morphologies are discussed in relation to the ability to form giant vesicles through the natural swelling method.
\end{abstract}

\end{frontmatter}

\section{Introduction}
All living cells on the earth maintain their lives by using a closed thin membrane with a size of 1-100 $\mu$m, the main constituent of which is phospholipid molecules. Currently, closed vesicular membranes made of phospholipids have been actively studied, and these have ranged in size from on the order of 10 nm to 10 $\mu$m. Among these vesicular membranes, giant vesicles, GV, which are larger than 1 $\mu$m, are thought to be suitable for use as a model system of living cells. \cite{1,2,3,27,28}. 

Various methods have been developed to prepare GVs such as the electro-formation or the solvent-spherule evaporation method \cite{4,5,6,7,25,26}. However, the use of such physical and chemical treatments inevitably has harmful effects on the model cell system, such as denaturation of proteins and the destruction of unstable biochemical species. On the other hand, under suitable conditions GVs are formed with mild treatment under natural swelling \cite{4}. Although this method has been frequently used, the mechanism of GV formation is not yet fully understood and the preparation method is still dependent on technical skill without a solid physico-chemical background \cite{6}. 

To investigate of the mechanism of GV formation by the natural swelling method, some studies with phase-contrast optical microscopy have shown that the myelin figures grew in the form of tubular fibrils when non-charged phospholipid dry films were hydrated \cite{4,8,9,10}. Reeves {\etal} considered the process of ``budding-off'' of the myelin figures as an intermediate state in GV formation \cite{4}. However, their experimental conditions were markedly different from the usual procedure of the natural swelling method, {\ie}, the phospholipid dry films were prepared filling the space between two parallel glass plates, and the amount of water was very small, while in the natural swelling a spot of phospholipid dry film is covered by excess water in a test tube. In addition, the ``budding-off'' process is not well verified by experiments and has not been understood theoretically. 

To solve the problems concerning the budding-off model, Lasic investigated the size distribution of vesicles hydrated from dry films on substrates with various topologies \cite{5}. They found that bilayered phospholipid flakes (BPF) peeled off from the support and their sizes depended on the topology of the substrate. However, this could not explain the mechanism of the natural swelling method. Since dry film has been thought to be aligned with slight defects on a smooth substrate, many defects should be formed in hydration process to make BPFs, where more energy is required to make defects compared with thermal fluctuation. In addition, the structures of phospholipid dry film composed of multi-stacked layers on a smooth substrate has not been directly observed, which might affect the formation of BPFs.

Recently, through the use of atomic force microscopy (AFM), the nm-scale structures of phospholipid films have been investigated \cite{11,12,13,14,20,21}. While these studies revealed the lateral structures of these films ({\eg}, domain formation, ripple structure), most observed only one or a few layers of membranes. Simonsen {\etal} showed the structure of spin-coated POPC dry film and hydrated dry film \cite{21}. In their study, the number of stacked layers was only 1-3 and they did not refer to the relation with the mechanism of GV formation. Leonenko {\etal} also performed AFM observation for only one layer of phospholipid on mica prepared by vesicle adsorption, and noted the temperature-dependence of the morphology \cite{24}. On the other hand, studies by X-ray scattering or neutron scattering on stacks of hundreds of layers on a solid substrate clarified only the average inter-layer structure, and did not aim at clarifying the mechanism of GV formation; the condition was far from that in the natural swelling method \cite{15,16,17}.

In this study, the micrometer-scale morphologies of phospholipid dry films composed of multi-layers on smooth substrate were observed as the initial state of GV formation by phase-contrast microscopy and AFM. All of the observations were carried out under the conditions that are commonly used to prepare GV with the natural swelling method. We also paid attention to the difference between the structures of the dry films in different phases; a disordered phase (liquid-crystalline $L_\alpha$ phase) and an ordered phase (gel $L_\beta{}'$ phase), above and below a main transition temperature $T_m$. To discuss the relation between the morphology of dry film and the efficiency of GV formation, hydrated products as the final state were observed by phase-contrast microscopy. The results revealed new aspects of GV formation.

\section{Materials and Method}
\subsection{Chemicals}
 1,2-Dioleoyl-sn-glycero-3-phosphocholine (DOPC) was obtained in powder form from SIGMA ALDRICH, 1,2-Dipalmitoyl-sn-glycero-3-phosphocholine (DPPC) was also obtained in powder form from Wako Pure Chemical Industries. The lipids were used without further purification. As a mother solution, these lipids were dissolved in an organic solvent composed of dehydrated chloroform and dehydrated methanol (2:1 v/v) (both of which were from NACALAI TESQUE). For complete dehydration, molecular sieves (NACALAI TESQUE) were mixed in the solution. The mother solution (10 mM) was stored at -30${}^\circ\! \rm{C}$ and used for all of the experiments after being allowed to stand at room temperature for five minutes. 
\subsection{Natural swelling method}
 Glass plates ($30\times40$ mm thickness 0.12-0.17 mm from MATSUNAMI GLASS) and test tubes ($9\times30$ mm from Maruemu Corporation) were treated with acetone. The mother solution (10 mM, 10 $\rm{\mu}$l) was dropped onto the glass plate or into the glass test tube, and the solvent was evaporated calmly in air (40-50\% humidity) at room temperature for over 15 minutes. To remove the solvent completely, the samples were placed under vacuum overnight. After this procedure, a 5-7mm-diameter spot of the phospholipid dry film remained on the glass plate or in the test tube. Under these conditions, over a hundred bilayers were stacked. The samples on glass plates were used for the observation of dry films and the samples in the test tubes were used to prepare GVs. When dry films in the test tube were hydrated with pure water (MilliQ, 100 $\rm{\mu}$l) at room temperature or at a temperature above the main transition point (from the gel phase to the liquid-crystalline phase) of phospholipid without agitation, GVs with diameters of 1-100 $\rm{\mu}$m swelled spontaneously (the suspension was at most 1 mM).

\subsection{Phase-contrast microscopy and atomic force microscopy (AFM)}
 The suspensions prepared by hydrating the phospholipid dry film or the phospholipid dry films on substrates were observed by phase-contrast microscopy (E600, Nikon) under green light. The images were captured with a CCD camera and recorded and analyzed with an image processor. To increase contrast in the images, 50 consecutive images were stacked and averaged. The phospholipid dry films on glass plates were also investigated by AFM (NVB-100, Olympus) with tapping mode. The images were analyzed and captured with a Nanoscope IIIa system (Digital Instruments). For modification, the images were flattened and plane-fitted to remove inclination of the glass plates. All experiments were carried out at room temperature

\section{Results}
\begin{figure}
\begin{center}
\includegraphics[width=.8\linewidth]{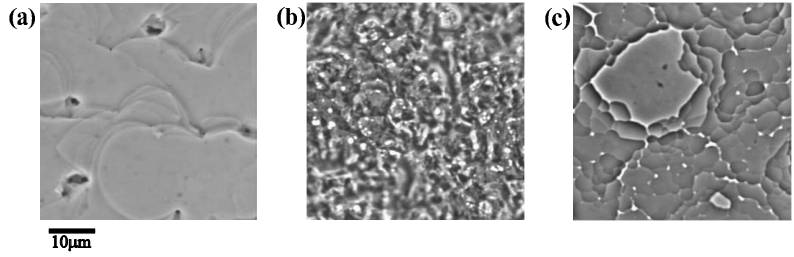}

\caption{Phase-contrast microscopic images of phospholipid dry films in air at room temperature. (a) DOPC dry film prepared in air at room temperature. The surface is flat and a stepped terrace-like morphology is observed. The size of the terrace is comparable to the size of GVs. (b) DPPC dry film prepared in air at room temperature. The surface seems to be rough, and no steps are observed. (c) DPPC dry film. The sample is prepared in air at room temperature and annealed to be in the liquid-crystalline phase (\textgreater 80${}^\circ\! \rm{C}$, \textgreater 12 h). After this procedure, the sample was observed at room temperature (in the gel phase). A step morphology is observed.
\label{film1}}
\end{center}
\end{figure}

\begin{figure}
\begin{center}
\includegraphics[width=.7\linewidth]{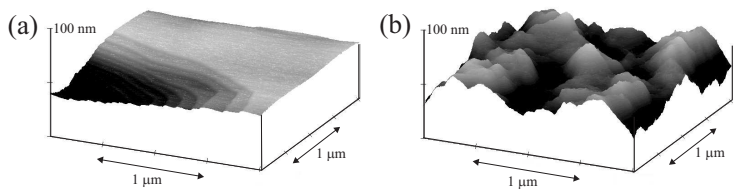}
\caption{3-D images of phospholipid dry films observed by AFM. (a) DOPC dry film showing a flat multilayered structure. The step height is about 4.5 nm, which corresponds to the thickness of a DOPC bilayer. (b) DPPC dry film exhibiting a random, rough surface.}
\label{film2}
\end{center}
\end{figure}

Figures \ref{film1}(a) and \ref{film1}(b) show the phase-contrast microscopic images of DOPC and DPPC dry films , both of which were prepared at room temperature (22${}^\circ\! \rm{C} \pm$2${}^\circ\! \rm{C}$). The surface of the DOPC film is smooth and flat, and exhibits a multi-layered terrace-like morphology. The size of each layer measured was 1-100 $\mu$m, which was comparable to the size of GVs obtained through natural swelling \cite{4}. In contrast, the surface of the DPPC film is rather rough. Figure \ref{film2} shows AFM images of DOPC and DPPC films. DOPC film were composed of several flat layers and the step height was about 4.5 nm, corresponding to the thickness of the DOPC bilayer in the liquid-crystalline $L_\alpha$ phase \cite{19}. In contrast, the DPPC dry film show a very rough surface.
 
Figure \ref{film1}(c) shows a phase-contrast microscopic image of a DPPC dry film observed at room temperature after being annealed at 80${}^\circ\! \rm{C}$ for over 12 hours. A flat multi-layered structure is observed, similar to the case of a DOPC dry film (Fig. \ref{film1}(a)).

\begin{figure}
\begin{center}
   \includegraphics[width=.75\linewidth]{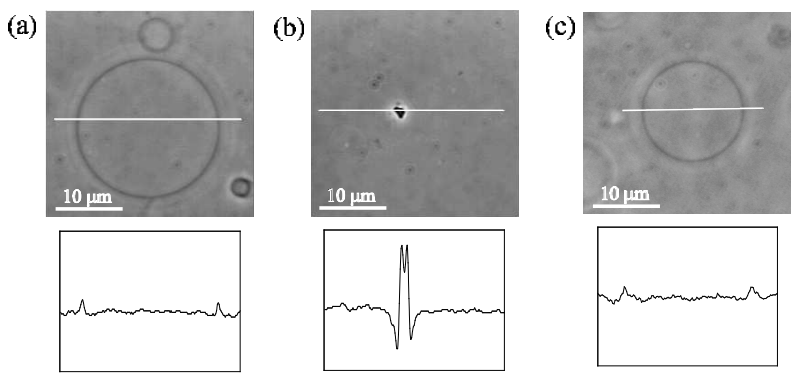}
   \caption{(Upper) Phase-contrast microscopic images of objects obtained by the natural swelling method (Lower) Cross-section profiles along the latitudinal lines. (a) Hydrated products of DOPC at room temperature. GVs are obtained effectively. (b) Phase-contrast images of DPPC. No GVs are generated. A few aggregates of phospholipid molecules are observed. (c) Phase-contrast image of DPPC hydrated at 50${}^\circ\! \rm{C}$. GVs are effectively obtained by hydrating DPPC film in the liquid-crystalline phase.} \label{liposome1}
\end{center}
\end{figure}

After 12 hours of hydration of these dry films with pure water at room temperature, the bulk aqueous solution was monitored by phase-contrast microscopy. Figures \ref{liposome1}(a) and \ref{liposome1}(b) show the typical products of DOPC and DPPC, respectively, swollen from dry films prepared at room temperature (corresponding to (a) and (b) in Figs. \ref{film1} and \ref{film2}, respectively). Differences were noted with regard to the efficiency of GV formation and the morphology of the suspension. Many GVs were obtained by hydrating the DOPC dry film, and the cross-section profile of DOPC has two small narrow peaks, which means that these products are vesicles made by thin-layered membranes (GV). The GVs ranged in the size from 1 to 100 $\mu$m and were polydisperse. In contrast, no GVs, but a few aggregates of phospholipid (\textless 1 $\mu$m), were obtained by hydrating the DPPC dry film. The cross-section profile of DPPC has a single large peak, which means that phospholipid molecules did not organize into giant vesicles but rather into random aggregates of phospholipid molecules. Notably, the DPPC film remained on the surface of the test tube after 12 hours of hydration. 

Figure \ref{liposome1}(c) shows the results of GV formation by hydrating DPPC dry film at 50${}^\circ\! \rm{C}$, {\ie}, which is above the main transition temperature. GVs were obtained effectively and the film did not remain on the surface of the test tube when the phospholipid was in the liquid crystalline phase. On the other hand, the DPPC dry film that was annealed at 80${}^\circ\! \rm{C}$ (Fig. \ref{film1}(c)) did not give either GVs or aggregations of phospholipid molecules when hydrated at room temperature.

\section{Discussion}
The present results show that liquid-crystalline and gel films of phospholipid exhibit smooth and rough layering. It is well known that DOPC is in the disordered, liquid-crystalline phase, and DPPC is in the ordered, gel phase, at room temperature under both dry and wet condition (in excess pure excess water, the main transition temperature $T_m$ of DOPC is about -20${}^\circ\! \rm{C}$ and that of DPPC is about 42${}^\circ\! \rm{C}$. In dry film, the $T_m$ of DOPC increases to 0${}^\circ\! \rm{C}$ and that of DPPC increases to 80${}^\circ\! \rm{C}$, respectively) \cite{18}. Phospholipid in the liquid-crystalline phase on a smooth substrate forms smooth, flat layers (Figs. \ref{film1}(a), \ref{film2}(a)). The flat layers of annealed DPPC indicates that this saturated phospholipid can form flat layers above the main transition temperature (Fig. \ref{film1}(c)). A disordered phospholipid film with a nm-scale length exhibits smooth, ordered layering on the scale of $\mu$m. In contrast, the morphology of a dry film in the gel phase depends on the process used for preparation. When organic solvents are evaporated below the $T_m$, the dry film exhibits rough layering (Fig. \ref{film2}(b)), while the well-aligned step morphology (Fig. \ref{film1}(c)) is formed when the dry film is prepared by cooling across the main transition temperature. In a related study, Perino-Gallice {\etal} indicated that wetting induces micrometer-scale changes in morphology \cite{20}. However, in the present study, similar morphologies were observed even in dry films and the difference in morphologies originates from the phases of phospholipid molecules. 

The hydration of phospholipid dry films gave simple results; in the liquid-crystalline phase, GVs were obtained effectively (Figs. \ref{liposome1}(a) and \ref{liposome1}(c)), while in the gel phase, GVs were not obtained at all (Fig. \ref{liposome1}(b)). Our results with DOPC showed that GVs could be formed effectively from the regular stacking of phospholipid layers on a smooth substrate, and suggested that the size of GVs was defined by the size of the phospholipid layers forming the terrace morphology. This tendency is consistent with the result reported by Lasic; the lipid layers of each terrace acts as a BPF. The highly aligned step morphology of a phospholipid dry film as an initial state is essential for the formation of GVs. Taking the result reported by Perino-Gallice {\etal} into account, the morphology of the phospholipid film was reconstructed and showed a well-aligned step morphology upon crossing the main transition temperature from the gel phase to the liquid-crystalline phase under wet conditions. GVs were formed effectively even when the morphology of the dry film was rough. Thus, GVs were effectively formed from the DPPC film at 50${}^\circ\! \rm{C}$.

The result that GVs were not formed effectively from the annealed DPPC dry film with the step morphology indicated that the large undulation motion of the membrane is also important for BPFs to peel off, since only steric force can act as a repulsive force for the non-charged phospholipid \cite{22}. This notion is closely related to the fact that the unbinding transition is induced only in the liquid-crystalline phase \cite{23}. 

\begin{figure}
\begin{center}
   \includegraphics[width=.75\linewidth]{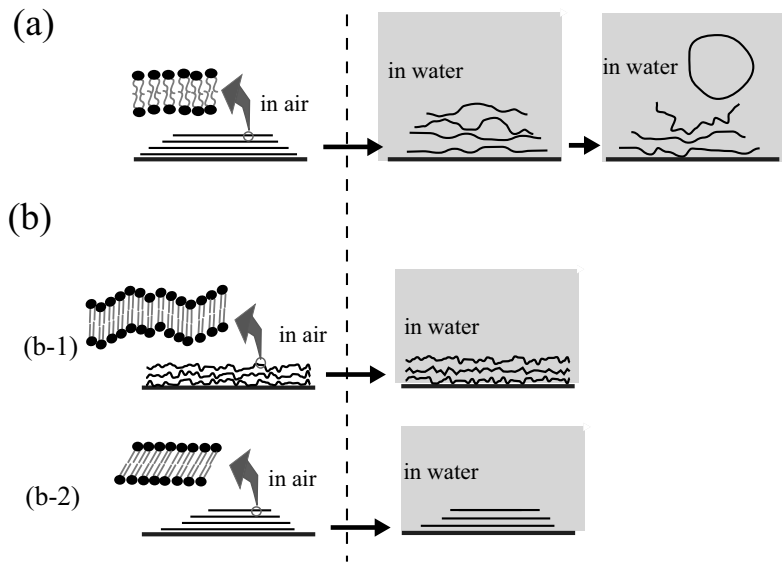}
   \caption{Schematic representation of the swelling of a phospholipid film. (Left) The structures of phospholipid dry films. (Right) Hydration process of the films. (a) Phospholipid in the disordered liquid-crystalline phase. The dry film is composed of regularly stacked layers with many steps. When the film is hydrated, membranes fluctuate greatly and are unbound from each other by steric repulsion. Finally, membranes peel off from the edges sequentially and their edges close to form vesicles. (b) Phospholipid in the ordered gel phase. The dry film is bumpy (b-1) or well aligned with steps (b-2). When the film is hydrated, it is difficult for membranes to separate and peel off because there are no edges and the membranes have less fluctuation.}
\label{mechanism}
\end{center}
\end{figure}

The mechanism of GV formation is depicted in the scheme in Fig. \ref{mechanism}. The phospholipid dry film in the disordered liquid-crystalline phase forms a regularly stacked step morphology like micrometer-scale terraces. When this film is hydrated, water penetrates between the layers and phospholipid membranes separate almost infinitely under thermal fluctuation. The steps in the multilayer serve to introduce aqueous solution into the interlayer region, and also promote sequential peeling off of the layers. The edge of a separated bilayer membrane adheres to form a vesicle. In contrast, when a film in the ordered gel phase is hydrated, phospholipid membranes cannot be unbound due to small undulation and most of the layers remain on the substrate without forming vesicles. 

\section{Conclusion}
In this study, we clarified that the morphologies of phospholipid dry films depend on the phase of the phospholipid, which is closely related to the efficiency of the formation of giant vesicles. Our results suggest that two conditions may be necessary for the effective formation of GVs. (i) the phospholipid film is well aligned and has many steps that form micrometer-scale terraces, and (ii) the membrane show adequate fluctuation in water.

\ack
{We are thankful to Dr. S-i. M. Nomura (Tokyo Medical and Dental University, Japan), N. L. Yamada (High Energy Accelerator Research Organization, Japan), and Mr. T. Hamada (Kyoto University, Japan) for their helpful advice. This work was supported in part by a Grant-in-Aid for the 21st century COE (Center for Diversity and Universality in Physics) from the Ministry of Education, Culture, Sports, Science, and Technology of Japan.}


\begin{thebibliography}{99}

\bibitem{1}
S-i.M. Nomura, K. Tsumoto, T. Hamada, K. Akiyoshi, Y. Nakatani, K. Yoshikawa, ChemBioChem 4 (2003) 1172
\bibitem{2}
P.-A. Monnard, J. Membr. Biol. 191 (2003) 87
\bibitem{3}
T. Baumgart, S.T. Hess, W.W. Webb, NATURE 425 (2003) 821
\bibitem{4}
J.P. Reeves, R.M. Dowben, J. Cell. Physiol 73 (1969) 49
\bibitem{5}
D.D. Lasic, J. Colloid Interfac. Sci. 124 (1988) 428
\bibitem{6}
D.D. Lasic, Biochem. J. 256 (1988) 1
\bibitem{7}
R.M. Watwe, J.R. Bellare, Current Science 68 (1995) 715
\bibitem{8}
D.M. Small, M.C. Bourg\`es, D.G. Dervichian, Biochim. Biophys. Acta 125 (1966) 563
\bibitem{9}
A. Saupe, J. Colloid Interface Sci. 58 (1977) 549
\bibitem{10}
W. Harbich, W. Helfrich, Chem. Phys. Lipids 36 (1984) 39
\bibitem{11}
J. Mou, J. Yang, C. Huang, Z. Shao, Biochemistry 33 (1994) 9981
\bibitem{12}
H.A. Rinia, B. Kruijiff, FEBS Lett. 504 (2001) 194
\bibitem{13}
A.F. Xie, R. Yamada, A.A. Gewirth, S. Granick, Phys. Rev. Lett. 89 (2002) 24610
\bibitem{14}
C. Leidy, T. Kaasgaard, J.H. Crowe, O.G. Mouritsen, K. J\o rgensen, Biophys. J. 83 (2002) 2625
\bibitem{15}
C. M\"unster, T. Salditt, M. Vogel, R. Siebrecht, J. Peisl, Europhys. Lett. 46 (1999) 486
\bibitem{16}
M. Vogel, C. M\"unster, W. Fenzl, T. Salditt, Phys. Rev. Lett. 84 (2000) 390
\bibitem{17}
G. Pabst, J. Katsaras, V.A. Raghunathan, Phys. Rev. Lett. 88 (2002) 128101
\bibitem{18}
K.L. Koster, Y.P. Lei, M. Anderson, S. Martin, G. Bryant, Biophys. J. 78 (2000) 1932
\bibitem{19}
J.F. Nagle, S. Tristram-Nagle, Biochim. Biophys. Acta 1469 (2000) 159
\bibitem{20}
L. Perino-Gallice, G. Fragneto, U. Mennicke, T. Salditt, F. Rieutord, Eur. Phys. J. E 8 (2002) 275
\bibitem{21}
A.C. Simonsen, A. Bagatolli, Langmuir 20 (2004) 9720
\bibitem{22}
W. Helfrich, Z. Naturforsch. 33a (1978) 305
\bibitem{23}
B. Pozo-Navas, V.A. Raghunathan, J. Katsaras, M. Rappolt, K. Lohner, G. Pabst, Phys. Rev. Lett 91 (2003) 028101
\bibitem{24}
Z.V. Leonenko, E. Finot, H. Ma, T.E.S. Dahms, T. Cramb, Biophys. J. 86 (2004) 3783
\bibitem{25}
M.I. Angelova, D.S. Dimitrov, Mol. Cryst. Liq. Cryst. 152 (1987) 89
\bibitem{26}
S. Kim, R. E. Jacobs, S.H. White, Biochim. Biophys. Acta 812 (1985) 793
\bibitem{27}
H. Hotani, F. Nomura, Y. Suzuki, Curr. Opin. Colloid Interface 4 (1999) 358
\bibitem{28}
S.L. Veatch, S.L. Keller, Biophys. J. 85 (2003) 3074
\end{thebibliography}
\end{document}